\documentclass[twocolumn, prl,showpacs]{revtex4}

\begin{document}

\title{Metaferroelectrics: Artificial Ferroelectricity in Metamaterials}

\author{Lasha Tkeshelashvili}

\affiliation{ Institut f\"ur Theoretische Festk\"orperphysik, Universit\"at Karlsruhe (TH), 76128 Karlsruhe, Germany }

\date{\today}

\begin{abstract}

Metamaterials are artificial periodic structures which represent effective homogeneous medium for electromagnetic fields. 
Here I show that there exists an important class of such composite systems, metaferroelectrics. They are properly 
designed metamaterials with non-ferroelectric constituents which, nevertheless, exhibit ferroelectric response. Furthermore, 
for relatively small samples the effect is anisotropic and strongly depends on the sample shape. Metaferroelectrics will 
help to understand deeper the physics of ferroelectricity and show enormous potential for applications in the field of 
nano-optoelectronics.
\end{abstract}



\maketitle

Metamaterials represent artificial periodic structures usually made of metallic nanoparticles. They exhibit a number of 
striking properties which lead to such unique effects as perfect lensing, electromagnetic cloaking, etc. \cite{busch2007}. 
The lattice constants of these systems are smaller than the wavelength of electromagnetic waves. Therefore, for the electromagnetic 
fields, metamaterials effectively represent a homogeneous medium. Perhaps this is the main reason why the role of periodicity was not 
fully appreciated for tailoring their electromagnetic properties. In reality, as is demonstrated below, the proper choice of lattice parameters is of paramount importance, and in certain cases, it determines qualitatively new phenomena such as artificial 
ferroelectricity.

Ferroelectrics have been a subject of intensive research for many years \cite{devonshire1954, ahn2004, dawber2005}. Nevertheless, the underlying physics of these materials remain less understood compared to their magnetic counterparts, ferromagnets \cite{akhiezer1968}. More than a half century ago Slater proposed that depolarization fields, stemming from the dipole-dipole interaction, may be responsible for appearance of spontaneous electric polarization \cite{slater1950}. However, it occurs that for existing ferroelectrics dipolar fields alone do not suffice \cite{slater1967}. Here, I show that in metamaterials with non-ferroelectric constituents the depolarization fields lead 
to the spontaneous polarization, and therefore, make possible to realize artificial ferroelectricity. 
Such metaferroelectrics will advance our understanding of physics of ferroelectricity and show a huge potential for possible applications~\cite{ahn2004}.

In particular, let us consider metamaterial which consists of cubic lattice of metallic nanospheres of radius $r$ embedded in a dielectric matrix with dielectric constant $\varepsilon_{\mbox{\tiny m}}$. Clearly, the lattice constant $d$ must obey $d\geq 2r$. To be specific the metal is assumed to be silver. Below the plasma frequency the dispersive properties of silver is best represented by the Drude dielectric function~\cite{slater1967}:
\begin{equation}
 \label{drude}
\varepsilon_{\mbox{\tiny Dr}}(\omega)=\varepsilon_\infty -\frac{\omega_{\mbox{\tiny p}}^2}{\omega(\omega-i\delta)},
\end{equation}
 where the plasma frequency $\omega_{\mbox{\tiny p}}$ is related to the density of free electrons $n_{\mbox{\tiny e}}$ as follows:
\begin{equation}
\label{plasmafr}
 \omega_{\mbox{\tiny p}}^2=\frac{n_{\mbox{\tiny e}}e^2}{\varepsilon_0m_{\mbox{\tiny eff}}},
\end{equation}
and $\delta$ is the damping constant. Further, $(-e)$ is the electron charge, $m_{\mbox{\tiny eff}}$ is the effective electron mass 
in the metal, and $\varepsilon_0$ is the permittivity of free space. 

The induced electric dipole moment ${\bf p}$ of an isolated silver sphere in the external electric field ${\bf E}$ 
is \cite{perenboom1981}:
\begin{equation}
 \label{dipolemoment}
{\bf p} = 4\pi r^3 \varepsilon_0 \varepsilon_{\mbox{\tiny m}}  
\frac{ \varepsilon_{\mbox{\tiny Dr}}-\varepsilon_{\mbox{\tiny m}} }{ \varepsilon_{\mbox{\tiny Dr}}+2\varepsilon_{\mbox{\tiny m}} } 
{\bf E}.
\end{equation}
This result is obtained for the static uniform electric field. However, it is still valid in the quasistatic limit when 
electromagnetic waves have wavelength much bigger than the sphere radius \cite{bohren1983}. Indeed, the comparison with exact Mie theory 
shows that for $3\mbox{nm}\leq r \leq 25\mbox{nm}$ the quasistatic approximation gives very accurate results. For the spheres with 
bigger radius quadrupole and higher order multipoles become important \cite{aroca2006}.

The dipole resonance frequency $\omega_{\mbox{\tiny R}}$, at which ${\bf p}$ would become infinite in the absence of losses, is 
$\mbox{Re}\{\varepsilon_{\mbox{\tiny Dr}}(\omega_{\mbox{\tiny R}})+2\varepsilon_{\mbox{\tiny m}}\}=0.$
This gives:
\begin{equation}
 \label{resfrequency}
\omega_{\mbox{\tiny R}}^2=\frac{\omega_{\mbox{\tiny p}}^2}{\varepsilon_\infty + 2 \varepsilon_{\mbox{\tiny m}}} -\delta^2 .
\end{equation}
It should be noted that, for $r\geq 3\mbox{nm}$ silver nanoparticles, the values of $\omega_{\mbox{\tiny R}}$ obtained on the bases of equation (\ref{resfrequency}) agrees very well with experimental data as well as with quantum mechanical calculations \cite{perenboom1981}. 
In the case of smaller spheres quantum size effects come into play and equation~(\ref{drude}) is no longer valid. 

The dynamical equation which correctly reproduces expression (\ref{dipolemoment}) reads~\cite{clippe1976, feng-qi1990}:
\begin{equation}
 \label{singlesphere}
\frac{d^2{\bf x}}{dt^2} + \gamma \frac{d{\bf x}}{dt} + \omega_{\mbox{\tiny R}}^2 {\bf x} = - \frac{q}{m} {\bf E},
\end{equation}
with ${\bf p} = - q {\bf x}$ \cite{slater1967}. $q$ is the absolute value of the oscillator charge $q=N_{\mbox{\tiny e}} e$, and 
$N_{\mbox{\tiny e}}$ is the total number of free electrons in the sphere $N_{\mbox{\tiny e}} = (4\pi r^3/3) n_{\mbox{\tiny e}}$. 
Further, ${\bf x}$ is the average electron displacement. The damping constant $\gamma$ and the mass of the oscillator $m$ are 
some functions of frequency \cite{clippe1976}. However, for our purpose only the values of these quantities in the static limit 
are important. At zero frequency $\gamma$ is an irrelevant parameter, and $m=g N_{\mbox{\tiny e}} m_{\mbox{\tiny eff}}$ with 
$ g= \omega_{\mbox{\tiny p}}^2 /(3 \varepsilon_{\mbox{\tiny m}} \omega_{\mbox{\tiny R}}^2) $.

Now, in an artificial lattice the local field acting on a sphere is ${\bf E} + (L/ \varepsilon_0) {\bf P} $, where $L$ is the 
depolarization factor \cite{slater1967}, and ${\bf P}$ is the polarization density. The equation of motion (\ref{singlesphere}) 
then becomes:
\begin{equation}
 \label{latticeeq}
\frac{d^2{\bf x}}{dt^2} + \gamma \frac{d{\bf x}}{dt} + \omega_{\mbox{\tiny R}}^2 {\bf x} = 
- \frac{q}{m} \left({\bf E} + \frac{L}{\varepsilon_0} {\bf P} \right).
\end{equation}
Taking into account that the unit cell volume is $d^3$, the polarization density can be written as ${\bf P}= (- q{\bf x})/d^3$. This allows to rewrite equation (\ref{latticeeq}) as:
\begin{equation}
 \label{finaleq}
\frac{d^2{\bf x}}{dt^2} + \gamma \frac{d{\bf x}}{dt} + \omega_{\mbox{\tiny eff}}^2 {\bf x} = 
- \frac{q}{m} {\bf E},
\end{equation}
with the effective frequency $\omega_{\mbox{\tiny eff}}$ given by
\begin{equation}
 \omega_{\mbox{\tiny eff}}^2 = \omega_{\mbox{\tiny R}}^2 - \frac{L q^2}{\varepsilon_0 m d^3 }.
\end{equation}
Assuming that $ {\bf E} \sim \exp{(i\omega t)}$ equation (\ref{finaleq}) gives:
\begin{equation}
 \label{finalpolarization}
{\bf P} = \frac{q^2/(m d^3)}{\omega_{\mbox{\tiny eff}}^2 -\omega^2 + i \gamma \omega} {\bf E}.
\end{equation}
The factor in front of ${\bf E}$ diverges at zero frequency when $ \omega_{\mbox{\tiny eff}}^2=0$. This means that arbitrarily small electric fields can induce nonzero electric dipole moment and the system becomes ferroelectric \cite{slater1967}. The condition 
 that the metamaterial is ferroelectric, $ \omega_{\mbox{\tiny eff}}^2=0$, can be cast in a more convenient form:
\begin{equation}
 \label{ferroelcond}
d_{\mbox{\tiny F}}^3=  4\pi L \varepsilon_{\mbox{\tiny m}} r^3.
\end{equation}
It is rather interesting that the lattice constant $d_{\mbox{\tiny F}}$ does not depend on the parameters of the Drude dispersion 
relation (\ref{drude}).
Let us assume that the background dielectric is silicon $\varepsilon_{\mbox{\tiny m}}=11.68$ \cite{salzberg1957}.
For the (infinite) cubic lattice $L=1/3$ \cite{slater1967}, and equation (\ref{ferroelcond}) gives $d_{\mbox{\tiny F}}=3.66r$.  
Here the following remark is in order. If the metallic nanospheres were point objects, and in the case of highly symmetric cubic 
lattice, the dipole approximation employed above would be correct for arbitrary $d_{\mbox{\tiny F}}$. However, the real spheres have 
a finite radius and the lattice constant must be bigger than certain $d_{\mbox{\tiny cr}}$. For $d \leq d_{\mbox{\tiny cr}} $ 
coupling between neighboring spheres excites quadrupole and other higher order modes and the dipole approximation becomes inaccurate. 
The detailed numerical studies (see e.g. \cite{ruppin1982, romero2006}) show that $d_{\mbox{\tiny cr}}=3.2r$. Since 
$ d_{\mbox{\tiny F}}> d_{\mbox{\tiny cr}}$ for the exemplary system considered here, the approximations involved are self-consistent 
and accurate.

Moreover, as was stressed above, the spheres behave as effective dipoles. The dipole fields decrease with the distance as $\sim 1/R^3$. On the other hand, the number of dipoles located at the distance $R$ from the given lattice site can be estimated as  $\sim R^2$. Therefore, if the sample size is $\sim R_s$, the field stemming from the surfice dipoles is proportional to $\sim 1/R_s$, which is a  slowly decreasing function. This means that, for relatively small systems, the depolarization fields depend on the sample shape \cite{akhiezer1968}. In general, they are inhomogeneous and anisotropic, and therefore, are represented as a space-dependent tensor. 

Nevertheless, for ellipsoids, the depolarization fields appear to be homogeneous. If the coordinate axes coincide with the principle axes of the ellipsoid, all non-diagonal elements of the depolarization tensor are identically zero. Furthermore, the following normalization relation $L_x+L_y+L_z=1$ holds, where $L_x$, $L_y$, and $L_z$ are the diagonal elements of the depolarization tensor~\cite{akhiezer1968}. 

In the case of prolate ellipsoid, 
\begin{displaymath}
 \frac{x^2}{a^2} + \frac{y^2+z^2}{b^2} = 1
\end{displaymath}
with the major and minor axes $a$ and $b$ respectively ($a>b$), we have:
\begin{eqnarray}
 L_x &=& \frac{1-e_{\mbox{\tiny PE}}^2}{2e_{\mbox{\tiny PE}}^3}
       \left( \ln \frac{ 1+e_{\mbox{\tiny PE}} }{ 1-e_{\mbox{\tiny PE}} } - 2e_{\mbox{\tiny PE}} \right), \nonumber\\
 L_y &=& L_z= \frac{1}{2}(1-L_x),
\end{eqnarray}
where the eccentricity $e_{\mbox{\tiny PE}} $  is given by $ e_{\mbox{\tiny PE}} = \sqrt{1-(b/a)^2} $. 

For a oblate ellipsoid, 
\begin{displaymath}
 \frac{x^2+y^2}{a^2} + \frac{z^2}{c^2} =1
\end{displaymath}
with the major and minor axes $a$ and $c$ respectively ($a>c$), the depolarization factors are:
\begin{eqnarray}
 L_x &=& L_y = \frac{1}{2}(1-L_z), \nonumber\\
 L_z &=& \frac{1+e_{\mbox{\tiny OE}}^2}{e_{\mbox{\tiny OE}}^3}
       \left( e_{\mbox{\tiny OE}}  - \arctan e_{\mbox{\tiny OE}} \right),
\end{eqnarray}
and the eccentricity $e_{\mbox{\tiny OE}} $ is $ e_{\mbox{\tiny OE}} = \sqrt{ (a/c)^2 -1}$.

Then, the equation of motion (\ref{finaleq}) takes the form:
\begin{equation}
 \label{finaleqgeneral}
\frac{d^2{\bf x}}{dt^2} + \gamma \frac{d{\bf x}}{dt} + \hat\omega_{\mbox{\tiny eff}}^2 {\bf x} = 
- \frac{q}{m} {\bf E},
\end{equation}
where the effective frequency tensor $\hat \omega_{\mbox{\tiny eff}}$ reads:
\begin{equation}
     \label{effomegatensor}
 \hat \omega_{\mbox{\tiny eff}}^2 = \left( \begin{array}{ccc}
\omega_{\mbox{\tiny R}}^2 - \frac{L_x q^2}{\varepsilon_0 m d^3 } & 0 & 0 \\
0 & \omega_{\mbox{\tiny R}}^2 - \frac{L_y q^2}{\varepsilon_0 m d^3 } & 0 \\
0 & 0 & \omega_{\mbox{\tiny R}}^2 - \frac{L_z q^2}{\varepsilon_0 m d^3 } 
        \end{array} \right).
\end{equation}
Note that ${\bf x}$ and ${\bf E}$ in equation (\ref{finaleqgeneral}) are represented by the column vectors. 

The system exhibits the ferroelectric response in a given direction if the corresponding diagonal element in equation (\ref{effomegatensor}) vanishes. Below some important particular cases are considered in more details:

(i) If a sample has the spherical shape $L_x=L_y=L_z=1/3$. Therefore, in this case the depolarization fields are the same as for 
the infinite lattice and $d_{\mbox{\tiny F}}^{\mbox{\tiny sp}}=3.66r$. The system response is isotropic and polarization density 
${\bf P}$ is a three-dimensional vector ${\bf P}=(P_x,P_y,P_z)$.

(ii) If a sample has the cylindrical or needle geometry which is oriented parallel to the $x$-axis $L_x=0$ and $L_y=L_z=1/2$.
In this limiting case equations (\ref{ferroelcond}) and (\ref{effomegatensor}) give  $d_{\mbox{\tiny F}}^{\mbox{\tiny cl}}=4.19r$. 
The system exhibits the easy-plane anisotropy and the polarization density is a planar vector ${\bf P}=(0,P_y,P_z)$. The ferroelectric
response is isotropic in the plane normal to the $x$-axis.

(iii) Finally, for a thin film or disk with the plane normal to the $z$-axis $L_x=L_y=0$ and $L_z=1$.
Thus, equations (\ref{ferroelcond}) and (\ref{effomegatensor}) suggest that $d_{\mbox{\tiny F}}^{\mbox{\tiny fl}}=5.27r$. The system 
shows easy-axis anisotropy, and therefore, the polarization density becomes a one-dimensional vector ${\bf P}=(0,0,P_z)$ with only 
one nonzero component.

In summary, here I have demonstrated that artificial ferroelectricity can be achieved in properly designed metamaterials. Most interestingly, the constituents of the composite system are assumed to be non-ferroelectric. In particular, the suggested structure consists of a cubic lattice of silver nanospheres embedded in a silicon matrix. Moreover, it should be stressed that for relatively small systems, due to the long-range nature of dipole-dipole  interaction, the corresponding lattice constant $d_{\mbox{\tiny F}}$ strongly depends on the sample shape and the effect becomes anisotropic. Metaferroelectrics show enormous potential for applications in the field of nano-optoelectronics.

I acknowledge support from the DFG-Forschungszentrum Center for Functional Nanostructures (CFN) 
at the Universit\"at Karlsruhe within Project No. A1.2 and from USA CRDF Grant No.~GEP2-2848-TB-06.

\end{document}